\documentclass[%
 reprint,
superscriptaddress,
 amsmath,amssymb,
 aps,prc
]{revtex4-1}
\usepackage{mathtools,amssymb,lipsum}
\usepackage{multirow}
\usepackage{dcolumn}
\usepackage{bm}
\usepackage[]{color}
\definecolor{mygreen}{rgb}{0.1, 0.6, 0.1}
\usepackage{amsmath}

\begin{document}

\preprint{APS/123-QED}

\title{The new MRTOF mass spectrograph following the ZeroDegree spectrometer \\at RIKEN's RIBF facility}

\author{M. Rosenbusch}
\affiliation{%
 Wako Nuclear Science Center (WNSC), Institute of Particle and Nuclear Studies (IPNS), High Energy Accelerator Research Organization (KEK), Wako, Saitama 351-0198, Japan\\
}%
\author{M. Wada}%
\affiliation{%
 Wako Nuclear Science Center (WNSC), Institute of Particle and Nuclear Studies (IPNS), High Energy Accelerator Research Organization (KEK), Wako, Saitama 351-0198, Japan\\
}
\author{S. Chen}%
\affiliation{%
 Department of Physics, The University of Hong Kong, Pokufulam, China\\
}%
\author{A. Takamine}%
\affiliation{%
 RIKEN Nishina Center for Accelerator-Based Science, Wako, Saitama 351-0198, Japan\\
}
\author{S. Iimura}
\affiliation{%
 RIKEN Nishina Center for Accelerator-Based Science, Wako, Saitama 351-0198, Japan\\
}%
\affiliation{%
 Department of Physics, Graduate School of Science, Osaka University, 1-1 Machikaneyama, Toyonaka, Osaka 560-0043, Japan\\
}%
\author{D. Hou}%
\affiliation{%
Institute of Modern Physics, Chinese Academy of Sciences, Lanzhou 730000, China\\
}%
\affiliation{%
University of Chinese Academy of Sciences, Beijing 100049, China\\
}%
\affiliation{%
School of Nuclear Science and Technology, Lanzhou University,  Lanzhou 730000, China\\
}%
\author{W. Xian}%
\affiliation{%
 Department of Physics, The University of Hong Kong, Pokufulam, China\\
}%
\author{S. Yan}%
\affiliation{%
 Institute of Mass Spectrometry and Atmospheric Environment, Jinan University, Guangzhou 510632, China\\
}%
\author{P. Schury}%
\affiliation{%
 Wako Nuclear Science Center (WNSC), Institute of Particle and Nuclear Studies (IPNS), High Energy Accelerator Research Organization (KEK), Wako, Saitama 351-0198, Japan\\
}%
\author{Y. Hirayama}%
\affiliation{%
 Wako Nuclear Science Center (WNSC), Institute of Particle and Nuclear Studies (IPNS), High Energy Accelerator Research Organization (KEK), Wako, Saitama 351-0198, Japan\\
}%
\author{Y. Ito}%
\affiliation{%
 Advanced Science Research Center, Japan Atomic Energy Agency, Ibaraki 319-1195, Japan\\
}%
\author{H. Ishiyama}
\affiliation{%
 RIKEN Nishina Center for Accelerator-Based Science, Wako, Saitama 351-0198, Japan\\
}%
\author{S. Kimura}%
\affiliation{%
 RIKEN Nishina Center for Accelerator-Based Science, Wako, Saitama 351-0198, Japan\\
}%
\author{T. Kojima}%
\affiliation{%
 RIKEN Nishina Center for Accelerator-Based Science, Wako, Saitama 351-0198, Japan\\
}%
\author{J. Lee}%
\affiliation{%
 Department of Physics, The University of Hong Kong, Pokufulam, China\\
}%
\author{J. Liu}%
\affiliation{%
 Department of Physics, The University of Hong Kong, Pokufulam, China\\
}%
\author{S. Michimasa}%
\affiliation{%
 Center of Nuclear Study (CNS), The University of Tokyo, Bunkyo 113-0033, Japan\\
}%
\author{H. Miyatake}%
\affiliation{%
 Wako Nuclear Science Center (WNSC), Institute of Particle and Nuclear Studies (IPNS), High Energy Accelerator Research Organization (KEK), Wako, Saitama 351-0198, Japan\\
}%
\author{J. Y. Moon}%
\affiliation{%
 Institute for Basic Science, 70, Yuseong-daero 1689-gil, Yusung-gu, Daejeon 305-811, Korea\\
}%
\author{M. Mukai}%
\affiliation{%
 RIKEN Nishina Center for Accelerator-Based Science, Wako, Saitama 351-0198, Japan\\
}%
\author{S. Nishimura}%
\affiliation{%
 RIKEN Nishina Center for Accelerator-Based Science, Wako, Saitama 351-0198, Japan\\
}%
\author{S. Naimi}%
\affiliation{%
 RIKEN Nishina Center for Accelerator-Based Science, Wako, Saitama 351-0198, Japan\\
}%
\author{T. Niwase}%
\affiliation{%
 Wako Nuclear Science Center (WNSC), Institute of Particle and Nuclear Studies (IPNS), High Energy Accelerator Research Organization (KEK), Wako, Saitama 351-0198, Japan\\
}%
\author{\\T. Sonoda}%
\affiliation{%
 RIKEN Nishina Center for Accelerator-Based Science, Wako, Saitama 351-0198, Japan\\
}%
\author{Y.X. Watanabe}%
\affiliation{%
 Wako Nuclear Science Center (WNSC), Institute of Particle and Nuclear Studies (IPNS), High Energy Accelerator Research Organization (KEK), Wako, Saitama 351-0198, Japan\\
}%
\author{H. Wollnik}%
\affiliation{%
 New Mexico State University, Las Cruces, NM 88001, USA\\
}%

\date{\today}

\begin{abstract}
A newly assembled multi-reflection time-of-flight mass spectrograph (MRTOF-MS) at RIKEN's RIBF facility became operational for the first time in spring 2020; further modifications and performance tests using stable ions were completed in early 2021. By using a pulsed-drift-tube technique to modify the ions' kinetic energy in a wide range, we directly characterize the dispersion function of the system for use in a new procedure for optimizing the voltages applied to the electrostatic mirrors. Thus far, a mass resolving power of $R_m > 1\,000\,000$ is reached within a total time-of-flight of only $12.5\,\mathrm{ms}$, making the spectrometer capable of studying short-lived nuclei possessing low-lying isomers. Detailed information about the setup and measurement procedure is reported, and an alternative in-MRTOF ion selection scheme to remove molecular contaminants in the absence of a dedicated deflection device is introduced. The setup underwent an initial on-line commissioning at the BigRIPS facility at the end of 2020, where more than 70 nuclear masses have been measured. A summary of the commissioning experiments and results from a test of mass accuracy will be presented. 
\end{abstract}

\maketitle


\section{\label{Intro}Introduction}
In the last decade, RIKEN's radioisotope Beam Factory (RIBF) \cite[]{Sakurai2010,Okuno2012,Motobayashi2012} enabled the discovery of more than 100 new radioisotopes (RI) at the extremes of the nuclear chart. The high yield of exotic nuclei has lead to the flourishing of experimental devices driven by a rich scientific program addressing major topics of nuclear physics. The scientific scope includes a wide range of r-process nucleosynthesis studies (see Sec. 3.2.1 in \cite{Kajino2019} and references therein), and studies of nuclear structure and fundamental properties of atomic nuclei (see \textit{e.g.} \cite[]{Steppenbeck2013,Taniuchi2019,Ahn2019,Huang2021,Kondo2016}). While until recently the scientific program has mainly been dedicated to reaction and decay spectroscopy, in the recent years it has begun to embrace high-precision mass spectrometry, and the facility now employs a magnetic storage ring \cite[]{Naimi2020,Nagae2021}, a B$\rho$ time-of-flight mass spectrometer \cite[]{Michimasa2013,Michimasa2020}, and multi-reflection time-of-flight mass spectrographs (MRTOF-MS) \cite[]{SCHURY201439,Ito2018}. Recently a new state-of-the-art MRTOF system referred to as the ZeroDegree MRTOF (ZD MRTOF) system was developed within the SLOWRI project, and has become operational at RIBF. This system has been coupled to a cryogenic helium-filled gas cell located behind the ZeroDegree (ZD) spectrometer \cite[]{Kubo2012} to slow down the reaction products delivered at relativistic energies to thermal equilibrium with the helium gas, and perform high-precision mass measurements of the RI at low kinetic energies.\par
Continuing advances in the performance of multi-reflection time-of-flight mass spectrographs play a vital role for fast and precise measurements of short-lived exotic nuclides produced at on-line radioactive ion beam (RIB) facilities worldwide. From its invention \cite[]{WOLLNIK1990267,WOLLNIK2003} onwards, every new step in the development of MRTOF-MS technology has given new opportunities, such as the resolution and separation of isobars \cite[]{PLA20084560,Wolf2012a,Wolf2013b,SCHURY2017a} and isomeric nuclear states \cite[]{DICKEL2015137,SanAndresUndChristine2019}, and the precise measurement of previously unknown nuclear masses (see \textit{e.g.} \cite[]{Wienholtz2013,Rosenbusch2015,Atanasov2015,Welker2017,Ito2018, Leistenschneider2018,HORNUNG2020,Schury2021,Mougeot2021,SBeck2021}). A strong point is the short duration of the measurement required to enable mass resolving powers of $R=\frac{m}{\Delta m}=\frac{TOF}{2\Delta TOF}>100\,000$, where $m$ and $TOF$ are the ion's mass and time-of-flight (TOF) while $\Delta m$ and $\Delta TOF$ are the full widths at half maximum (FWHM) of the mass and TOF spectral peaks, respectively. This duration is typically on the order of ten milliseconds or less, while occasionally extending to a few tens of milliseconds. Presently various research groups around the world are taking part in the development of MRTOF systems at RIB facilities, including RIKEN/RIBF (Japan) \cite[]{ISHIDA2004,ISHIDA2005,SCHURY201439}, CERN/ISOLDE (Switzerland) \cite[]{Wienholtz2013,Wolf2012a} , GSI (Germany) \cite[]{PLA20084560,PLA2013457}, TRIUMF (Canada) \cite[]{Jesh2017,Leistenschneider2018}, Argonne (USA) \cite[]{HIRSH2016229}, Notre Dame University (USA) \cite[]{SCHULTZ2016251}, GANIL (France) \cite[]{CHAUVEAU2016}, IMP (China) \cite[]{WANG2020179}, and JYFL (Finland). Apart from these applications, MRTOF technology has advanced to other research fields including cluster physics \cite[]{Knauer2019}, and also neutrino physics where the focus is on unambiguous identifications of rare events \cite[]{Murray2019}.\par
The basic principle of the MRTOF-MS is the repetitive reflection of ions between two (typically concentric) electrostatic ion mirrors with the aim to focus ions onto a detector with a TOF distribution as narrow as possible after their long flight path. To this end the ion-optical aberrations must be minimized, whereby one of the crucial ingredients for MRTOF systems is the distribution of mirror bias potentials forming a characteristic shape of the electric potential, and the geometry of the electrodes. On the one hand, these mirror bias potentials must be optimized to keep radial ion-optical aberrations small, where one important development in this context was the introduction of strongly negative voltages at certain positions in the system (negative ion-optical lenses) \cite[]{WOLLNIK2003}, which also serve for radial confinement in general. On the other hand, the potential distribution defines the TOF-energy dispersion function (see \textit{e.g.} \cite[]{Wolf2012}), \textit{i.e.} the relationship between the kinetic energy of an ion and its final time-of-flight: ${TOF}(E_\mathrm{kin})$. This function has been taken into account for the optimization of mirror electrodes in this work, and the correction of ${TOF}(E_\mathrm{kin})$ yielded a useful initialization to reach high resolving powers.\par
When applying the MRTOF-MS to rare ions, an additional possible challenge can derive from the existence of numerous stable molecular ions -- some possibly having a much greater intensity than the RI -- in the analysis ensemble. A typical case of such occurs when using a helium gas cell to thermalize relativistic ion beams, where molecules become ionized either directly by interaction with the incoming accelerator beam or in charge exchange with the ionized helium. When these stable molecular ions have a sufficiently similar mass-to-charge ratio $A/q$ to that of the radioactive ions being analyzed, they can be useful as mass references. However, when the $A/q$ differ sufficiently from that of the radioactive ions then they will reach the detector after performing a different number of laps inside the MRTOF-MS, which can result in a contaminant ion's TOF signal interfering with that of the desired ion and hindering the unambiguous assignment of any spectral peak. To solve such problems, a growing trend not only for nuclear physics but also for analytical chemistry \cite[]{Johnson2019} is to perform in-MRTOF selection of ions, \textit{i.e.} removing unwanted ions while they reflect within the MRTOF-MS. The most prominent method is the application of a deflection device in the MRTOF-MS, first applied to studies of molecular ions \cite[]{Toker2009}, later adopted for the research on exotic nuclei \cite[]{Dickel2015}, and more recently studied with an MRTOF setup for atomic clusters \cite[]{Fischer2018}. An alternative approach has been reported from analytical chemistry, where the mirror endcaps of an MRTOF-MS have been repetitively opened shortly for selected ion ejection \cite[]{Johnson2019}. Similarly, a scheme using a single voltage pulse for the central drift tube of an MRTOF-MS has been used for the removal of isobaric contaminants \cite[]{WIENHOLTZ2017285}. We introduce an alternative method for wideband in-MRTOF ion selection developed for the absence of a deflection device. It is based on repeatedly pulsing the mirror electrodes close to the central drift tube of the device, and does not require a pulsing of very sensitive potentials as applied to the endcaps.\par
In this work, the new setup along with the applied voltages will be presented in detail. This will be followed by a discussion of in-MRTOF ion selection, and optimization of the mirror electrode voltages by measuring TOF-energy functions using a pulsed drift tube to change the ions' mean kinetic energy. Finally, we discuss the on-line commissioning of the system, which was done in parallel to in-beam $\gamma$-ray experiments (see \textit{e.g.} \cite[]{Steppenbeck2013}) at the end of 2020.
\section{\label{Setup}Experimental setup}
The new device has a similar design and operation to the setup reported in \cite[]{SCHURY201419}. Herein we present the electrode configuration and the applied voltages. Furthermore, an alternative in-MRTOF cleaning method, allowing for selective ion removal without the need of an in-MRTOF deflector, will be discussed. The vacuum system and the experimental timing sequence will be explained as well.
\subsection{Electrode Setup and applied Voltages}
\label{sub:electrodes}
\begin{figure*}[]
\includegraphics[width=0.95\textwidth]{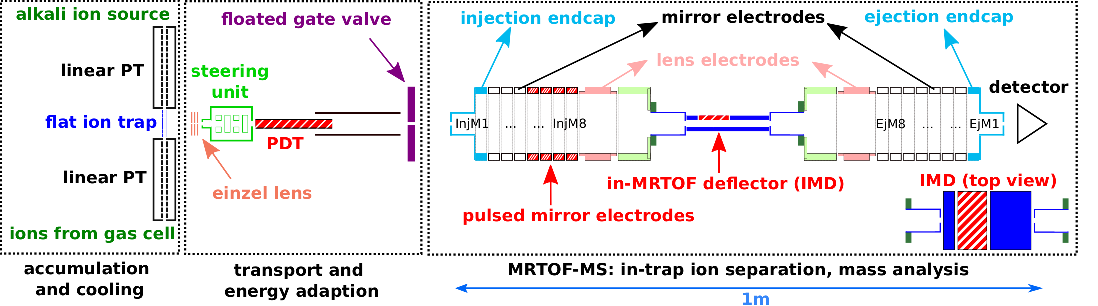}
	\caption{\label{fig:electrodes} Electrode configuration of the present MRTOF setup, where $\mathrm{InjM1...M8}$ refer to the eight mirror electrodes at the injection side and $\mathrm{EjM1...M8}$ to the ones at the ejection (detector) side. Red color with white stripes is chosen for electrodes pulsed for manipulation of the ions' energy, and for ion deflection. Other abbreviations: PT for Paul trap, PDT for pulsed drift tube (see also text).}
\end{figure*}
The new mass-measurement setup is dedicated to high-precision nuclear mass measurements at the BigRIPS on-line facility of RIKEN in Japan. The new spectrometer has been coupled to a novel gas cell \cite[]{Takamine2019,Iimura2021} equipped with radiofrequency carpets (see \textit{e.g.} \cite[]{WADA2003570,SAVARD2003,WEISSMAN2004,WADA2013,DICKEL2016,SUMITHRARACHCHI2020}) wherein relativistic radioactive ions are stopped, extracted and transported to the MRTOF system. The present electrode design of the MRTOF-MS setup is shown in Fig.~\ref{fig:electrodes} and \cite[]{ROSENBUSCH2020}.\par
For off-line studies and optimization of experimental parameters, a thermal ion source delivering alkali ions is used. It also serves as a source of reference ions for high-precision mass measurements when the device is operated on-line. Ions from the gas cell and from the thermal ion source are simultaneously injected as continuous beams into their adjacent segmented linear Paul traps (PT) and are then transferred, cycle-by-cycle from each side, to a planar PT referred to as flat ion trap (FT) located between the two linear PTs. It is used for dedicated cooling (without further accumulation of ions) and for properly locating the ion cloud to prepare the perpendicular ejection towards the MRTOF device through a small exit aperture of $0.7\,\mathrm{mm}$ diameter. The cycle-by-cycle injection of ions into the central trap from either source, gas cell or reference ion source, enables the concomitant referencing technique \cite[]{SCHURY201840}.\par 
Upon ejection from the FT the ions are accelerated in a transfer section, which connects the MRTOF-MS with the ion traps. An accelerating einzel lens composed of three plates, each having an aperture of $3\,\mathrm{mm}$, is used to provide spatial focusing while accelerating the ions to $E_\mathrm{kin} \approx 1.4\,\mathrm{keV}/q$. Beyond the third plate the ions pass a double steerer unit, which is biased with the potential of the third plate of the einzel lens. The deflector section is followed by a pulsed drift tube (PDT) ($15\,\mathrm{cm}$ in length and $1.5\,\mathrm{cm}$ in diameter) the bias of which is rapidly increased from $V_\mathrm{PDT}^\mathrm{low} = -1.4\,\mathrm{kV}$ to about $V_\mathrm{PDT}^\mathrm{high} = +1.1\,\mathrm{kV}$ while the ions are inside the field-free space at the inner of the PDT. As the potential of the central drift section in the MRTOF-MS has an outsized influence on the time-of-flight, we have chosen to keep it a ground potential to enforce its stability. The PDT increases the ion's energy to allow them to pass the mirror electrodes without the need to float the ion trap on high-voltage. As will be discussed later, it also provides an opportunity to directly measure the energy dispersion function of the MRTOF mirrors.\par
The ions exit the PDT and enter a static drift tube of larger diameter, which is floated to the same voltage $V_\mathrm{drift}^\mathrm{static} = V_\mathrm{PDT}^\mathrm{high}$, and move for a distance of $\approx 30\,\mathrm{cm}$ until reaching the entrance of the MRTOF device. The ions are axially confined inside the MRTOF-MS by fast switching the bias applied to the injection endcap (first mirror electrode), where the ions pass with a kinetic energy close to $2.5\,\mathrm{keV}$ (high-voltage switching by in-house built fast transistor switches). They are reflected back and forth for a predetermined number of laps (for a specific $A/q$) before being released by fast switching of the bias applied to the ejection endcap (last mirror electrode). After exiting the MRTOF-MS, ions travel to a fast ion impact detector (ETP 14DM572) and the time duration between their ejection from the flat ion trap and their impact on the detector is digitized with a $100\,\mathrm{ps}$ precision multi-hit time-to-digital converter (MCS6a from FAST ComTec).\par
The presently applied approximate voltages for the ion-transfer and MRTOF-MS are given in Tab.~\ref{transfer-voltages} together with the relative sensitivity coefficients $C_\mathrm{i} = \frac{\delta {TOF}_i / {TOF}}{ \delta V_i / V}$ of the mirror electrodes, where $\delta {TOF}_i$ is the change in the TOF induced by the change in voltage $\delta V_i$.\par
\begin{table}
	\caption{List of approximate voltages used for ion transfer from the flat ion trap to the MRTOF device, and for ion reflections. Electrode name, voltage, sensitivity coefficient (only for voltages used in reflection mode), and description.}
\begin{tabular}{c | c | c | l}
	\label{transfer-voltages}
	name & voltage & $ C_\mathrm{i} $ & description \\
	\hline
	$ V_\mathrm{FT}$ 		& $-10\,\mathrm{V}$ 	&--& center of flat ion trap \\
	$ V_\mathrm{Eje+}$ 		& $+190\,\mathrm{V}$ 	&--& FT positive ejection\\
	$ V_\mathrm{Eje-}$ 		& $-210\,\mathrm{V}$ 	&--& FT negative ejection \\
	$ V_\mathrm{EL1}$ 		& $-300\,\mathrm{V}$ 	&--& first einzel lens plate \\
	$ V_\mathrm{EL2}$ 		& $-320\,\mathrm{V}$ 	&--& second einzel lens plate \\
	$ V_\mathrm{EL3}$ 		& $-1400\,\mathrm{V}$ 	&--& third einzel lens plate \\
	$ V_\mathrm{Def}$ 		& $-1400\,\mathrm{V}$ 	&--& deflector section floating \\
	$ V_\mathrm{PDT}^\mathrm{low}$ 	& $-1400\,\mathrm{V}$ 	&--& pulsed drift tube low \\
	$ V_\mathrm{PDT}^\mathrm{high}$ 	& $+1100\,\mathrm{V}$ 	&--& pulsed drift tube high \\
	$ V_\mathrm{drift}^\mathrm{static}$ 		& $+1100\,\mathrm{V}$ 	&--& static drift tube \\
	$ V_\mathrm{valve}$ 		& $+1100\,\mathrm{V}$ 	&--& floated gate valve \\
	$ V_\mathrm{InjM1}^\mathrm{open}$ 	& $+1230\,\mathrm{V}$ 	&--& open mirror 1 injection side\\
	$ V_\mathrm{InjM1}^\mathrm{closed}$ 	& $+3080\,\mathrm{V}$ 	&$-24.49\,\%$& closed mirror 1 injection side\\
	$ V_\mathrm{InjM2}$ 		& $+2220\,\mathrm{V}$ 	&$-3.17\,\%$& mirror 2 injection side\\
	$ V_\mathrm{InjM3}$ 		& $+2540\,\mathrm{V}$ 	&$+0.65\,\%$& mirror 3 injection side\\
	$ V_\mathrm{InjM4...8}$ 	& $0\,\mathrm{V}$ 	&--& grounded mirrors\\
	$ V_\mathrm{InjL}$ 		& $0\,\mathrm{V}$ 	&--& lens injection side\\
	$ V_\mathrm{drift}^\mathrm{MRTOF}$ 	& $0\,\mathrm{V}$ 	&--& MRTOF drift section\\
	$ V_\mathrm{EjL}$ 		& $-4890\,\mathrm{V}$ 	&$-3.63\,\%$& lens ejection side\\
	$ V_\mathrm{EjM7,8}$ 		& $+900\,\mathrm{V}$ 	&$+1.28\,\%$& mirror 7,8 ejection side\\
	$ V_\mathrm{EjM6}$ 		& $+1030\,\mathrm{V}$ 	&$+1.98\,\%$& mirror 6 ejection side\\
	$ V_\mathrm{EjM5}$ 		& $+2950\,\mathrm{V}$ 	&$+8.00\,\%$& mirror 5 ejection side\\
	$ V_\mathrm{EjM4}$ 		& $+1320\,\mathrm{V}$ 	&$+2.54\,\%$& mirror 4 ejection side\\
	$ V_\mathrm{EjM3}$ 		& $+2603\,\mathrm{V}$ 	&$-5.01\,\%$& mirror 3 ejection side\\
	$ V_\mathrm{EjM2}$ 		& $+2260\,\mathrm{V}$ 	&$-8.92\,\%$& mirror 2 ejection side\\
	$ V_\mathrm{EjM1}^\mathrm{closed}$ 	& $+3500\,\mathrm{V}$ 	&$-19.37\,\%$& closed mirror 1 ejection side\\
	$ V_\mathrm{EjM1}^\mathrm{open}$ 	& $+500\,\mathrm{V}$ 	&--& open mirror 1 ejection side\\
	\hline
\end{tabular}
\end{table}
The MRTOF reflection chamber consists of a pair of electrostatic mirrors separated by a grounded central drift tube. Each mirror comprises eight shorter mirror electrodes and one longer electrode intended to serve as an ion optical lens (see Fig.~\ref{fig:electrodes} and \cite[]{ROSENBUSCH2020} for geometric details). The ion impact detector is located about $5\,\mathrm{mm}$ away from the ejection endcap electrode. \par
The voltage configuration in the MRTOF-MS is asymmetric whereby the three outermost mirror electrodes on the injection side form a steep potential well, while the five other electrodes and lens are nominally grounded. The biases applied to the ejection mirror form a shallow electric potential whose shape results in production of a time-focus at a desired number of laps in the reflection chamber. All mirror voltages are applied through low-pass RC filters with $1-2\,\mathrm{s}$ time constant for the innermost mirror electrodes ($1\,\mathrm{M \Omega}$, $1-2\,\mathrm{\mu F}$), $5\,\mathrm{s}$ time constant for $V_\mathrm{EjM2,3,4,5}$ ($5\,\mathrm{M \Omega}$, $1\,\mathrm{\mu F}$), and $40\,\mathrm{s}$ time constant for the endcap electrodes $V_\mathrm{InjM1}^\mathrm{closed}$ and $V_\mathrm{EjM1}^\mathrm{closed}$ ($5\,\mathrm{M \Omega}$, $8\,\mathrm{\mu F}$).\par
The application of low-pass filters improves the stabilization of voltages, but also brings drawbacks for the usage of switched mirrors. During continuous switching of voltages, electric charges are transferred between the two applied voltage sources (\textit{i.e.} open and closed state of an ion mirror). As there are resistors separating each buffer capacitor from its voltage supply, the voltage at the electrode differs from that of the supply. There is a complicated interplay between the RC filters and the switch. With each switching cycle, the filter's capacitor is partially drained to charge the capacitance of the electrode, and must be recharged. The current by which the power supply recharges the filter's capacitor results in a voltage drop across the filter's resistor. The steady state voltage drop across the filter's resistor thus depends on the switching-cycle frequency. As such, the timing system has been prepared to allow for an interruption-free operation. A further active voltage stabilization \cite[]{SCHURY201439,Wienholtz2020,Fischer2021,Schury2019} is anticipated at the ZD MRTOF setup for the future.\par
Using the sensitivity coefficients, the mass resolving power $R_m$ which can be achieved in the approximation of independent random voltage fluctuations is given by
\begin{equation}
	R_{m} = \frac{1}{2} \frac{TOF}{\Delta TOF + TOF \cdot \sqrt{\sum_i \left(C_i \cdot \left(\delta V / V \right)_i \right)^2}}\,,
\end{equation}
where $\left(\delta V / V \right)_i$ is the relative voltage noise for each electrode and $\Delta TOF$ the width of the ions' TOF distribution at the detector. The dominant contribution for the TOF drifts comes from the outermost mirror electrodes and yields a requirement of an average voltage fluctuation of $\overline{\delta V / V} \approx 1.4\cdot 10^{-6}$ to reach $R_m \approx 10^6$ if the width of the ion distribution is not considered ($\Delta TOF \rightarrow 0$), and $\overline{\delta V / V} < 10^{-6}$ for realistic values of $\Delta TOF$. This denotes the minimum relative stability which must persist for a time interval within which a reference spectrum of sufficient statistics is obtained to accommodate software-based drift correction.\par
\subsection{\label{ITS} In-MRTOF ion selection and the alternative method PMiMS}
As previously noted, a challenge for the new setup comes from contaminant molecules with different mass-to-charge ratios. These molecules are ionized in the gas cell by the incoming high-energy ion beam and charge exchange with the ionized helium gas. This leads to the dominant presence of ions with different lap times in the MRTOF-MS, and the subsequent detection of TOF peaks with unknown number of laps overlapping with the ions of interest in the same spectrum. Although the simultaneous injection of as many as a few hundred ions is not generally expected to cause non-negligible space charge effects when being distributed across multiple mass numbers, the identification of rare ions in such contaminated TOF spectra is very challenging. Hence, a removal of all non-isobaric contaminant ions is essential for the success of on-line experiments.\par
\begin{figure}[b]
\includegraphics[width=0.45\textwidth]{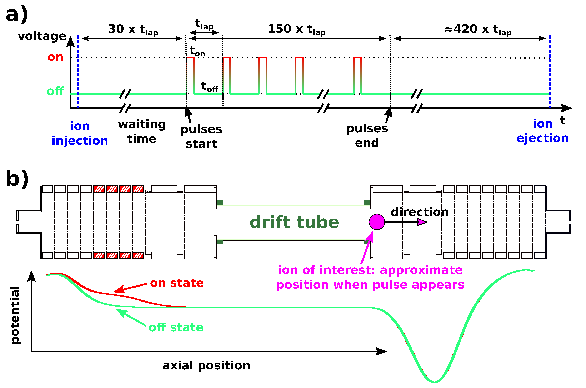}
	\caption{\label{PMiMS} Illustration of the PMiMS scheme. a) Pulse timing structure: pulses are synchronized to the motion of the wanted ion species during the ion trapping period, where $t_\mathrm{lap}$ is the lap time of the ions of interest and $t_\mathrm{on}$ ($t_\mathrm{off}$) is the on (off) state duration of the pulse in each lap. b) Potential shape for on/off state of the $600\,\mathrm{V}$ pulse and position of the wanted ion species at the start of the $t_\mathrm{on}$ time.}
\end{figure}
During the first on-line commissioning of the setup in 2020 (see results in Sec.~\ref{Online}), a pulsed-mirror in-MRTOF selection (PMiMS) scheme has been developed as an alternative means of in-MRTOF ion selection (see Fig.~\ref{PMiMS}). Initially, the measurement of exotic species was hindered by the large amount of contaminant molecular ions delivered from the gas cell, and at the time of the experiment a deflector unit had not yet been installed in the MRTOF-MS. To nonetheless succeed in performing the measurement, periodic high-amplitude voltage pulses were applied to the four (otherwise grounded) mirror electrodes indicated in Fig.~\ref{fig:electrodes} and \ref{PMiMS}b. The sudden change of the potential acts similar to a pulsed drift tube and causes a change of the kinetic energy of ions present in the affected region, with the intention to make their trajectories unstable. The energy can be increased or decreased depending on the polarity of the voltage transition.\par
The pulses were produced using a fast transistor switch altering the potential between the electrical ground of the system and a high-voltage power supply set to $\approx 600\,\mathrm{V}$. As the electric noise propagating through the system can induce a disturbance on other electrodes, the chosen voltage was a compromise between effectively ejecting the unwanted ions and maintaining high accuracy for the mass measurement of isobaric chains. The pulses were synchronized with the reflection period of the ions of interest, and the phase was chosen so that the desired ions were located in a non-affected region when the pulse was present.\par
\begin{figure}[t]
\includegraphics[width=0.45\textwidth]{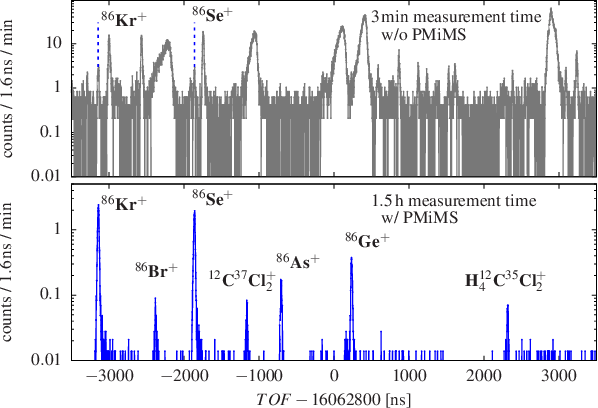}
	\caption{\label{cleaning} On-line application of in-MRTOF cleaning using electric pulses applied to four mirror electrodes. Two spectra of $A/q=86$ ions delivered from the BigRIPS facility have been recorded. In the top the cleaning method is not applied and molecular contaminants of other masses performing different numbers of laps in the MRTOF-MS are present throughout the TOF spectrum. The different width of the ion peaks comes from the difference of the lap number as the system is operated in dispersive mode (time focus after a fixed number of laps). In the bottom, the purified spectrum is shown, where only species with $A/q = 86$ are remaining.}
\end{figure}

The technique relies on providing a sufficient energy change to the unwanted ions, so that they either overcome the potential barrier of the endcap electrodes, or touch an electrode due to an unstable trajectory. The energy gain (or loss) depends on the physical location of an unwanted ion at the time of the PMiMS pulse, and thereby more than one pulse may be required for a given ion species. It was found that an optimal performance was reached by using 150 pulses in equally many subsequent laps (total number of laps for wanted ions species was mostly 600 during the experiments), with a duration of $3-5\,\mathrm{\mu s}$ for each pulse. Before starting the PMiMS pulses after ion injection into the MRTOF-MS, a waiting period of about 30 laps was applied to provide time for spatial ion separation. This purification scheme was successful and the resulting TOF spectra were sufficiently free from non-isobaric contaminants to provide high-precision mass analysis of the exotic radioactive ions. In Fig.~\ref{cleaning} a pair of TOF spectra, with and without the PMiMS applied, is shown for $A/q = 86$ isobars. In the upper panel of Fig.~\ref{cleaning}, only two species could be identified; the spectral peaks corresponding to the other $A/q=86$ ions are lost in the contaminant-ion spectral peaks. The remaining species have successfully been identified with PMiMS technique applied, as shown in the lower panel of the figure.\par
After the first experiments using PMiMS, a new prototype of in-MRTOF deflector (IMD) with a planar geometry has been installed in the central drift region of the ZD MRTOF system (shown in Fig.~\ref{fig:electrodes}), and is presently used. Two stainless steel plates are placed in a distance of $1\,\mathrm{cm}$, where one of them is separated to host a pulsed $5\,\mathrm{cm}$ long electrode enabling the deflection of ions. The lateral width of all plates is $10\,\mathrm{cm}$ to provide sufficient electric shielding of the area. By the choice of plate distance and size, a single hit with a dipole field of only $30\,\mathrm{V/cm}$ is necessary to cause an ejection of unwanted ions within the same lap, \textit{i.e} before they can exit the IMD section. This allows for a targeted selective ejection at chosen lap numbers and, thus, also for a selective maintaining of ions with different $A/q$ while ejecting all other ions.\par
The PMiMS method described herein was a solution according to the circumstances at the time being, and enabled the performance of on-line experiments. Thus, it is worth to be reported as a possible alternative, and this technique may be useful for other research groups using the MRTOF technique.
\subsection{Vacuum setup}
\label{sub:vacuum}
\begin{figure}[ht]
\includegraphics[width=0.47\textwidth]{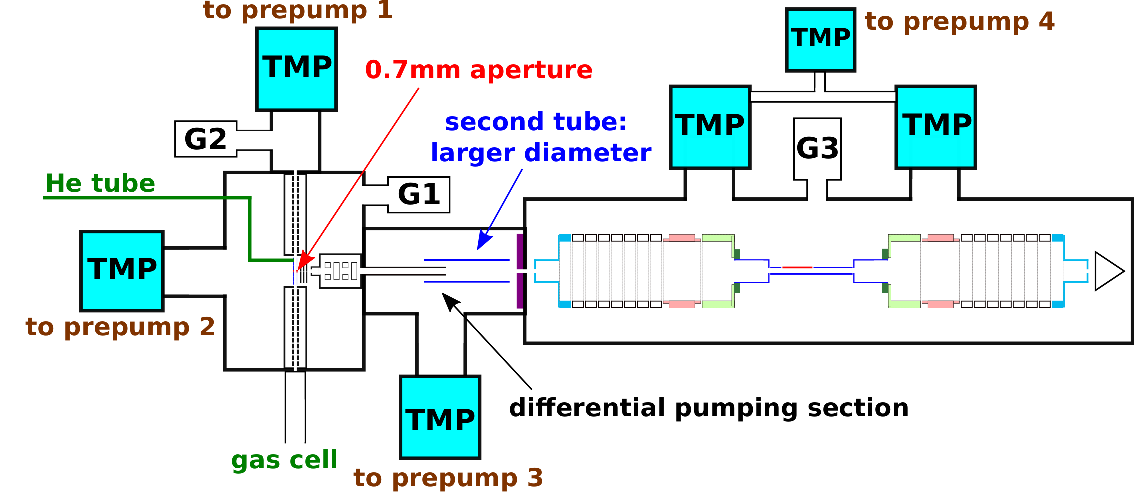}
  \caption{\label{fig:vacuum} Sketch of the vacuum system with tree differential pumping stages to reduce the helium load in the MRTOF-MS. The symbols G1,G2,G3 denote vacuum gauges.}
\end{figure}

The vacuum system (see Fig.~\ref{fig:vacuum}) has been optimized to allow for sufficient helium gas pressure in the Paul traps while preserving ultra-high vacuum inside the MRTOF chamber, while also keeping the length of the transfer section moderately short. To this end the linear Paul traps are tightly encapsulated by thin metal tubes, which serve to mount them within the ion-trap suite. The tubes are mounted inside the holder of the flat ion trap and provide a seal against the outer trap chamber using viton o-rings. The circuit boards with the electrode structure of the flat ion trap are affixed to the holder using Caramaseal vacuum-compatible epoxy to construct a leak-tight system. Helium gas is injected directly into the flat ion trap using a PTFE tube connection on the trap holder, which also supplies the linear Paul traps. The only leakage from the trap region is due to the necessary $0.7\,\mathrm{mm}$ aperture for ion ejection. The inner volume of the traps and the surrounding vacuum chamber are pumped by separate turbomolecular pumps (TMP), providing the first two essential pumping stages (Edwards STP-451, and Edwards STP-H451C).\par
The steerer unit following the flat ion trap is installed in a separate sealed housing with an entrance aperture of $6\,\mathrm{mm}$ diameter, and at the exit the pulsed drift tube is mounted. Both units are pumped by the transfer section units provide a further pumping barrier between ion trap and MRTOF-MS. The transfer section is separately pumped by a TMP (SEIKO SEKI STP-301) with independent pre-pump and the adjacent tube to the pulsed drift tube has a larger diameter to improve the gas flow to the TMP. The final pumping barrier is a $6\,\mathrm{mm}$ aperture at the entrance of the MRTOF chamber. The chamber of the mass analyzer is pumped by two TMPs in parallel (Edwards STP-451), where a small TMP (Edwards NEXT85D) is installed on the forelines to improve the compression ratio and allow for a better removal of residual helium from the trap chamber. In the present configuration the nominal pressure in the mass analyzer is $P=3\cdot 10^{-6}\,\mathrm{Pa}$ when the trap is optimally pressurized.\par
\begin{figure}[t]
	
\includegraphics[width=0.45\textwidth]{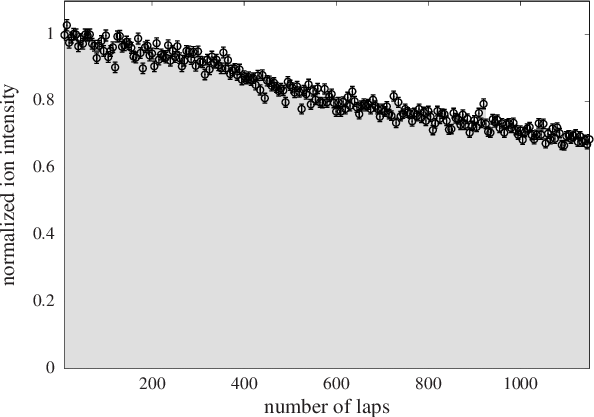}
	\caption{\label{counts-over-laps} Normalized ion intensity of $^{39}$K$^{+}$ ions as a function of the number of laps performed in the MRTOF-MS. At the highest investigated lap number the ions' flight path is $2.2\,\mathrm{km}$ while each lap has a flight path of approximately $1.8\,\mathrm{m}$. The corresponding TOF for this $A/q$ is on the order of $20\,\mathrm{ms}$.}
\end{figure}

Despite the additional efforts, the collisions of ions with the residual helium gas in the MRTOF-MS causes some losses with increasing flight path. As shown in Fig.~\ref{counts-over-laps}, the transmission of singly-charged $^{39}$K$^{+}$ ions over the course of up to 1200 laps is measured to be about $70\,\mathrm{\%}$, where the flight path is on the order of $2\,\mathrm{km}$ and the time of flight $21\,\mathrm{ms}$. Typical operation is performed using 500 to 700 laps, where $\approx80\,\mathrm{\%}$ of the potassium ions are transmitted. The estimated mean free path is $\approx 5\,\mathrm{km}$.\par
We note that not only the vacuum in the MRTOF-MS chamber but also the cleanliness of the trapping system is crucial to avoid ion losses. Off-line tests show that chemically non-reactive alkali ions can be processed and transported to the MRTOF-MS efficiently. However, for other ion species, reactions with contaminant atoms and molecules, charge exchange, or neutralization could cause significant losses. These mechanism of ion losses turned out to be of major importance for the performance of gas cell, while losses due to the (well optimized) trapping and transport play a minor role (see Sec.~\ref{Online} for system efficiency in on-line operation).
\subsection{Measurement sequence}
\label{sub:timing_sequence}
As described in \cite[]{SCHURY201840}, to achieve a very high duty cycle both linear Paul traps can be filled with ions independently (and simultaneously) from each source, reference ion source and gas cell. Both of the linear Paul traps transfer ions to the central flat ion trap in an alternating fashion. After a cooling period in the flat ion trap, the ions are ejected and transferred to the MRTOF system. Except for the brief transfer to the flat ion trap, both linear Paul traps are continuously in accumulation and cooling mode. While one linear trap injects ions into the central trap, the potential of the other linear trap enables accumulation and trapping, where the potential of the segment closest to the flat ion trap is a reflecting well. This alternate injection enables the concomitant referencing method \cite[]{Ito2018,SCHURY201840} which allows for interleaved measurements of reference ions and those from the gas cell. The present cycle time is $50\,\mathrm{ms}$, which allows for excellent compensation of TOF drifts caused by voltage fluctuations and thermal expansion of the mechanical structure.\par
\begin{figure}[t]
\includegraphics[width=0.5\textwidth]{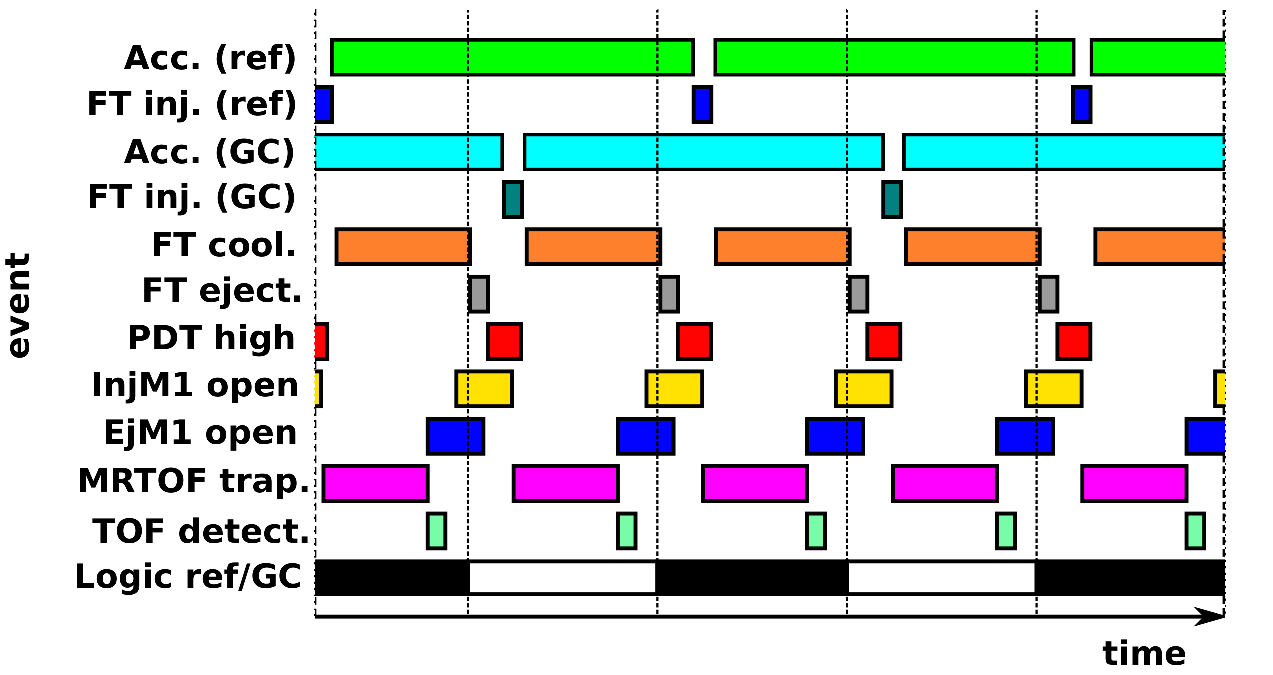}
	\caption{\label{fig:timing} Interleaved event scheme of the measurement sequence. \textbf{Acc. (ref):} Accumulation and cooling of ions in reference Paul trap ($\approx 40\,\mathrm{ms}$). \textbf{FT inj. (ref):} Injection of ions from the reference trap into flat ion trap (few microseconds). \textbf{Acc. (GC):} Accumulation of ions in the Paul trap adjacent to the gas cell ($\approx 40\,\mathrm{ms}$). \textbf{FT inj. (GC):} Injection of ions from gas cell side Paul trap into flat ion trap (few microseconds). \textbf{FT cool.:} Cooling of ions in flat ion trap ($2-7\,\mathrm{ms}$). \textbf{FT eject.:} Ejection of ions towards the MRTOF section. \textbf{PDT high:} Pulsed drift tube at $ V_\mathrm{PDT}^\mathrm{high}$ ($ V_\mathrm{PDT}^\mathrm{low}$ otherwise). \textbf{InjM1 open:} Injection endcap at open state (closed otherwise). \textbf{EjM1 open:} Ejection endcap open state. \textbf{MRTOF trap.:} Symbolic for reflection period in MRTOF system. \textbf{TOF detect.:} TOF detection (typically $10-20\,\mathrm{ms}$ TOF). \textbf{Logic ref/GC:} Logic tag signal for data acquisition. See \cite[]{SCHURY201840} for an alternative illustration.}
\end{figure}

The full measurement sequence consists of two sub-cycles (presently of $25\,\mathrm{ms}$ duration) including the full procedure from ion accumulation to detection. The interleaved timing sequence is shown in fig.~\ref{fig:timing}, where ions from the gas cell (thermal ion source) are analyzed in the even (odd) sub-cycles. The timing patterns for both ion species are similar: cooling in the flat ion trap, ejection from this trap, PDT switching, raising the injection mirror potential after the ions' passage, and lowering the ejection mirror potential when the ions have performed the desired number of laps in the MRTOF-MS. The timings for the pulsed drift tube and both mirror endcaps are mass dependent, which requires a pre-calculated timing pattern in each sub-cycle. The trigger signals for the high-voltage transistor switches are provided by an in-house built FPGA sequencer. The free-running timing system is built around a Spartan 6 FPGA and provides both TTL and NIM signals with $1\,\mathrm{ns}$ precision.
\section{\label{Optimization}Ion mirror optimization and off-line performance}
The goal of the optimization of the mirror potentials is to achieve a narrow ion time-of-flight distribution at the detector. This requires that (besides other ion-optical aberrations) the TOF difference coming from the different kinetic energies of the ions must be minimized, \textit{i.e.} achieving the same time-of-flight for all energies imprinted on the ions by the extraction field. The extraction field in the preparation trap forms the correlated part of the ions' position-energy distribution, which is converted into a TOF distribution at the detector (see Fig. 18 and 19 of \cite[]{Wollnik1999} for analog spatial imaging in sector magnet spectrometers). The function ${TOF}(E_\mathrm{kin})$ is nonlinear in general and depends on all ion-optical elements in the system. The major effort for the optimization is to find an optimal potential shape for the ion mirrors, allowing for a high-quality time focus. The achievable limit comes from the ions' turnaround time due to the finite thermal energy distribution, being the uncorrelated part of the position-energy profile. Finding the best conditions is a multi-parameter problem, and requires meaningful procedures to succeed. \par
To that end, after an initial approximate tuning of the mirror potentials, the system's dispersion function is determined; the time-of-flight as a function of energy using the PDT to scan through the ion energy, similar to the measurements shown in \cite[]{Wolf2013a,WIENHOLTZ2017285}. Ions ($^{39}$K$^{+}$) were accelerated into the pulsed drift tube at different electric potentials $V_\mathrm{PDT}^\mathrm{low}$, whereas the post-pulse voltage $V_\mathrm{PDT}^\mathrm{high}$ was not changed. The ions were then stored in the MRTOF-MS for $690\,\mathrm{laps}$ and ejected towards the detector. The dispersion function ${TOF}(E_\mathrm{kin})$ can only be approximately measured as ${TOF}(\overline{E_\mathrm{kin}})$, containing the average TOF produced by the internal energy distribution of the ion cloud. We note that probably the most accurate approximation of the dispersion function would be achieved by directly varying the trap potential instead of an element in some distance to the trap. However, the potential $V_\mathrm{PDT}^\mathrm{low}$ has been used instead to measure ${TOF}(\overline{E_\mathrm{kin}})$ due to present technical limitations of the trap potentials.\par
\begin{figure}[t]
\includegraphics[width=0.45\textwidth]{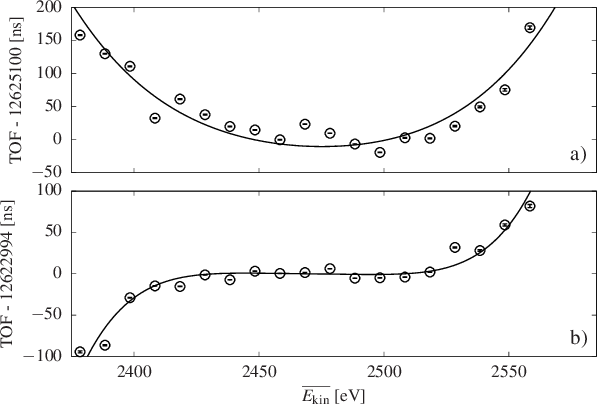}
	\caption{\label{fig:ToF-E-1} a) Time-of-flight of $^{39}$K$^{+}$ ions as a function of the mean kinetic energy varied by the electric potential $V_\mathrm{PDT}^\mathrm{low}$ of the negative (entrance) voltage of the pulsed drift tube for a detuned mirror potential allowing isochronous behavior only for a small energy interval. b) Same measurement for well-tuned mirror potentials showing a significantly larger isochronous energy interval.}
\end{figure}
In Figure~\ref{fig:ToF-E-1} a) an example of the ${TOF}(\overline{E_\mathrm{kin}})$ function yielding only moderate resolving powers of $R_m \approx 200,000$ is shown. Across the $100\,\mathrm{eV}$ wide central span of the parabolic profile -- ranging from $\overline{E_\mathrm{kin}} = 2425\,\mathrm{eV}$ to $2525\,\mathrm{eV}$ -- the TOF varies by at about $50\,\mathrm{ns}$. The goal was to reduce the TOF variance for a larger interval of kinetic energies. In this example a shorter TOF for lower kinetic energies was desired. This was accomplished by raising the potential around the inner mirror electrodes, \textit{i.e.} close to the drift tube, which has an influence on the position at which the ions with lower kinetic energy are reflected.\par
The curve in Fig.~\ref{fig:ToF-E-1} b) is the result of the same measurement after raising the potential applied to the 5$^{th}$ mirror electrode on the ejection side $V_\mathrm{EjM5}$ by $30\,\mathrm{V}$, followed by a refocusing procedure using the strongly negative potential $V_\mathrm{EjL}$. The shape is significantly different and the region of flat response has been expanded. Within $\Delta E_\mathrm{kin} = 100\,\mathrm{eV}$ the TOF varies only by about $20\,\mathrm{ns}$. This setting results in a higher mass resolving power with already $R_m > 500\,000$, and proved to be a useful initial configuration for the later fine tuning of the mirror voltages.\par
The fine tuning was done by smaller changes of the potentials followed by a recovery of the focus length of the system (TOF after which a focus is achieved). When the MRTOF-MS is already tuned so that a first-order time focus is present at the chosen number of laps, then a significant change of a single voltage will detune this focus. In this situation, it cannot be known a priori if this change would enable to find a better or worse time focus. It is thus necessary to either change the number of laps in order to search for the new focus, or to use another electrode not involved in the sensitive region of ion reflection (\textit{e.g.} central drift tube, negative lens electrode, or front-end electrodes between trap and MRTOF-MS) to re-adjust the focus length of the system.\par
To this end, after changing a mirror voltage in the ion-reflection region, the negative lens electrode on the ejection side was used to move the time focus back to the initial number of laps. Although this lens electrode influences the lateral focusing of ions and is an element primarily responsible for ion confinement, the response is weak enough to allow for larger voltage changes without harming the confinement. Furthermore, the change in bias applied to such an electrode (far lower voltage than ion-reflection region) has only marginal influence on the ${TOF}(\overline{E_\mathrm{kin}})$ profile. In this way, the potentials $V_\mathrm{EjM1c,2,3,4,5}$ have been changed in $2-10\,\mathrm{V}$ steps, to chose a new shape for the electric potential in the reflection region, followed by the refocusing procedure before the resolving power has been determined for each setting.\par
\begin{figure}[b]
\includegraphics[width=0.48\textwidth]{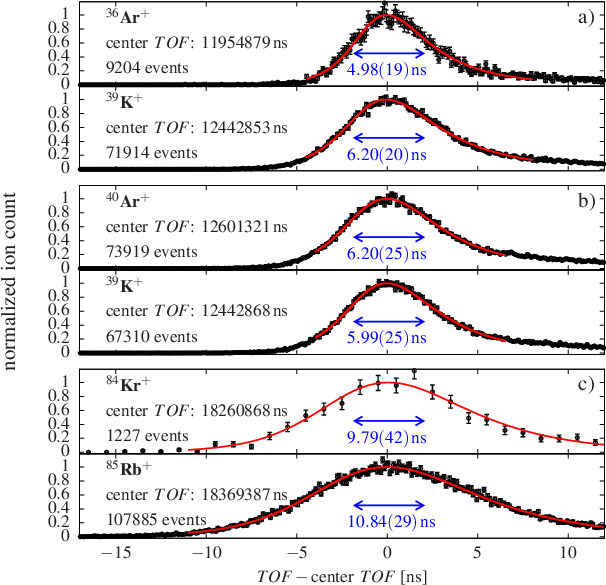}
	\caption{\label{fig:spectra} a) Time-of-flight spectra for $690\,\mathrm{laps}$ measured in pairs (analyte/reference) for one hour accumulation time. a) $^{36}$Ar$^{+}$ from the gas cell with $^{39}$K$^{+}$ ions as reference yielding $R_m \approx 1\,200\,000$. b) $^{40}$Ar$^{+}$ from the gas cell with $^{39}$K$^{+}$ ions as reference yielding $R_m \approx 1\,000\,000$. c) $^{84}$Kr$^{+}$ from the gas cell with $^{85}$Rb$^{+}$ ions as reference yielding $R_m \approx 900\,000$ (see Tab.~\ref{MRP} for the full number set). The peak width refers to the FWHM.}
\end{figure}
As a benchmark for the efforts, stable atomic ions $^{36,40}$Ar$^{+}$ and $^{84}$Kr$^{+}$ have been delivered from the gas cell. Traces of these noble-gas ions which are present as low-level contaminants in the helium gas were ionized by an $\alpha$-emitting source isolated from the gas cell by a Mylar window thin enough to transmit alpha particles. At the same time, reference alkali ions have been produced from the thermal ion source at the other side of the trap chamber, where $^{39,41}$K$^{+}$ were used as references for ions having $A/q \approx 40$, and $^{85}$Rb$^{+}$ for ions having $A/q \approx 85$. Time-of-flight spectra have been measured using the concomitant referencing method, and were analyzed following a software-based drift correction. The drift correction was performed by tracking the drift of the alkali reference ion's TOF peak. Independent of the rate of ions delivered from the gas cell, the rate of reference ions was always tuned to be about one ion per cycle in average. Each pair of spectra (analyte spectrum, reference spectrum) was recorded for one hour to achieve reasonably high statistics and to obtain a robust estimate of the mass resolving power. The resulting TOF spectra (shown in Fig.~\ref{fig:spectra}) have been fitted by a Johnson SU distribution \cite[]{Johnson1949a,Johnson1949b} suited to fit symmetric as well as skewed peak shapes.\par 
The results are shown in Tab.~\ref{MRP}. According to the recent works from other research groups discussing high resolving powers (see \textit{e.g.} \cite[]{YAVOR20181,Mardor2021}), and to the present knowledge of the authors, this is a new record for MRTOF mass spectrometry. This achievement will facilitate the resolution of low-lying isomers in future MRTOF mass measurements. A resolving power of $R_m \approx 10^6$ enables the identification of nuclear isomers (at FWHM level) with excitation energies of, for example, $38\,\mathrm{keV}$ for $A/q \approx 40$, $93\,\mathrm{keV}$ for $A/q \approx 100$, and $186\,\mathrm{keV}$ for $A/q \approx 200$. This feature, in combination with decay-correlated mass spectroscopy (see \textit{e.g.}  \cite[]{Niwase2020}), will allow unambiguous identification of the state ordering, an important detail which is presently lacking in many cases.\par
In Penning-trap experiments, the phase-imaging ion-cyclotron-resonance (PI-ICR) technique performed since 2013 \cite[]{Eliseev2013,BLOCK2016265}, enabled a 40-fold increased resolving power as compared to the previously used TOF-ICR technique \cite[]{KONIG199595}, and mass resolving powers of $R_M = 10^7$ have been reported with measurement durations on the order of a second. As the mass resolving power of Penning traps scales linearly with the duration of the measurement, the MRTOF-MS technique becomes competitive for measurement durations as low as $20\,\mathrm{ms}$, considering the performance reported herein.
\begin{table}
	\caption{Ion species, resolving power $R_m$ achieved after one hour of accumulation time using software drift correction for the reference ions, spectrum identifier (a-c as used in Fig.~\ref{fig:spectra}), and comments.}
\begin{tabular}{c | c | c | c}
	\label{MRP}
	ion species & $ R_m $ & spectrum & comment\\
	\hline
	$^{36}$Ar$^+$ & $1.20(4)\cdot 10^{6}$ & \multirow{2}*{a} & analyte \\
	\cline{1-2}
	\cline{4-4}
	$^{39}$K$^+$ & $1.00(3)\cdot 10^{6}$ &  & reference \\
	\hline
	\hline
	$^{40}$Ar$^+$ & $1.02(5)\cdot 10^{6}$ & \multirow{2}*{b} & analyte \\
	\cline{1-2}
	\cline{4-4}
	$^{39}$K$^+$ & $1.04(4)\cdot 10^{6}$ &  & reference \\
	\hline
	\hline
	$^{84}$Kr$^+$ & $0.93(4)\cdot 10^{6}$ & \multirow{2}*{c} & analyte \\
	\cline{1-2}
	\cline{4-4}
	$^{85}$Rb$^+$ & $0.85(3)\cdot 10^{6}$ &  & reference \\
	\hline
\end{tabular}
\end{table}
\section{On-line Experiments}
\label{Online}
\begin{figure}[b]
\begin{center}
\includegraphics[width=0.48\textwidth, clip]{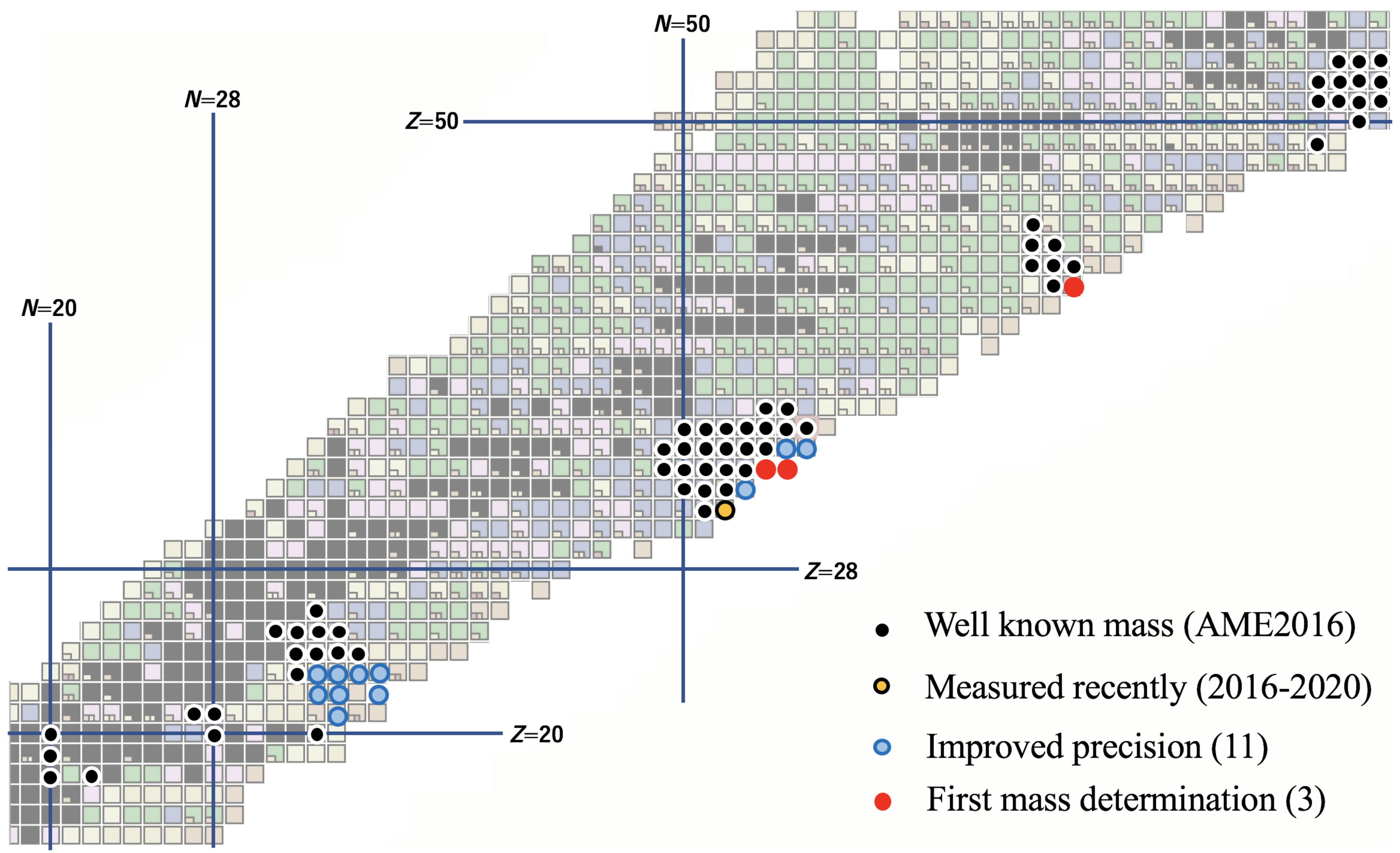}
\end{center}
	\caption{Nuclear mass measurements during parasitic operation and the results by the ZD MRTOF system (adapted from \cite[]{Marco2021}). Background color code illustrates the measured precision from AME (see Nucleus-win for reference).}
\label{fig:Chart}
\end{figure}

In 2020 the MRTOF system was coupled to the new cryogenic gas cell and was relocated to a position behind the ZeroDegree spectrometer. The system was ready to operate on-line just before the start of the 2020 winter campaign of in-beam $\gamma$-ray experiments \cite[]{Doornenbal2012,Doornenbal2016} of the new HiCARI project \cite[]{Wimmer2021,Doornenbal2018}, which enabled the first on-line commissioning of the new ZD MRTOF setup in parasitic operation. The RI were produced by in-flight fission of a uranium beam delivered by RIKEN's superconducting ring cyclotron (SRC) accelerator with a beam energy of $345\,\mathrm{MeV/u}$. The production took place in beryllium targets of $5-11\,\mathrm{mm}$ thickness, depending on the isotopes requested. The desired reaction products were selected by the BigRIPS separator and passed to the HiCARI detector array and the ZeroDegree spectrometer before being implanted in the gas cell. In such parasitic operation, the gas cell plays the role nominally performed by a lead beam dump. Energy degraders were employed to reduce the beam energy to enable a stopping position inside the helium gas. The stopped reaction products were extracted mostly as singly-charged atomic ions and as molecular sidebands upon chemical reactions in the helium gas. The ions were transported to the MRTOF setup, where after applying the PMiMS scheme introduced in Sec.~\ref{ITS} the atomic masses of the various implanted nuclides were determined with high precision and accuracy. The 2020 HiCARI campaign lasted throughout the month November and December comprising a total of seven individual experiments, all of which used the gas cell as a beam dump. During these commissioning runs, mass measurements covering four different regions in the nuclide chart have been carried out (see Fig~\ref{fig:Chart}).\par
\begin{figure}[t]
\begin{center}
\includegraphics[width=0.48\textwidth, clip]{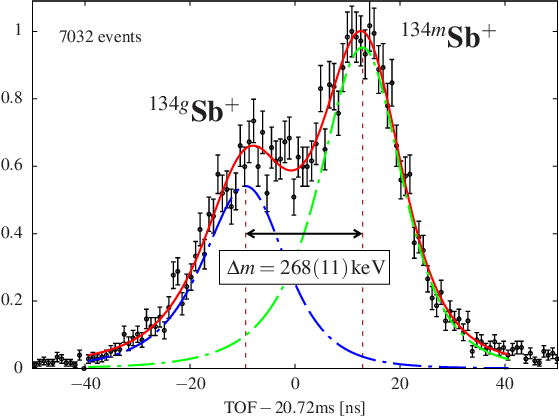}
\end{center}
	\caption{Identification of nuclear ground state and isomeric state of $^{134}$Sb$^{+}$ with a resolving power of $R_m = 550\,000$ achieved prior to the most recent optimization (adapted from \cite[]{Wenduo2021}).}
\label{fig:Sb}
\end{figure}

As the commissioning experiments spanned several weeks, the degree of contamination in the gas cell varied across the measurement campaign. Consequently, the efficiency varied based on the chemical nature of a given atomic nucleus. The most abundant observed molecular sideband for $^{55}$Sc ($^{55}$ScOH$^+$) had a total efficiency of $e<10^{-4}$, while the less reactive $^{85}$As$^{+}$ and $^{137}$Te$^{+}$ were found to have total efficiencies of $0.16\,\mathrm{\%}$ and $1.3\,\mathrm{\%}$, respectively \cite[]{Iimura2021}.
The major mission of the commissioning run was the test of components using on-line beam. Nevertheless, the commissioning was highly successful for the study of nuclear masses with more than 70 atomic masses measured during the campaign. Around neutron-rich Ti and V isotopes, our results include isotopes from Sc to Fe with $^{55}$Sc, $^{58}$Ti, and $^{59}$V being the most exotic ones measured in this region, improving upon the precision of masses recently measured using the TOF-B$\rho$ method at the NSCL \cite{Meisel2020} and RIBF \cite{Michimasa2020} at the forefront of nuclear mass studies in this region. In the neutron-rich region above Ni, nuclides have been studied spanning from Ga to Kr with $^{84}$Ga, $^{86}$Ge, $^{89}$As, and $^{91}$Se being the most exotic species. Among those isotopes, $^{88,89}$As have been measured for the first time while three other mass values provide a significant improvement over previously performed measurements. Another region of isotopes successfully addressed spans from Mo to Rh including the first mass measurement of  $^{112}$Mo. In the fourth region addressed near $^{132}$Sn the separation of the ground state and the isomeric states of $^{134g,m}$Sb has been demonstrated reaching a resolving power of $R_m = 5.5\cdot 10^5$ at that time (see Fig.~\ref{fig:Sb}). We note that this measurement occured prior to the new tuning and measurements reported in Sec.~\ref{Optimization}. The excitation energy of the isomer in $^{134}$Sb was measured to be $\Delta E = 268(11)\,\mathrm{keV}$ which is in agreement with $\Delta E = 279(1)\,\mathrm{keV}$ given in the literature \cite[]{JAIN20151}. In total, atomic masses have of three isotopes have been measured for the first time, and atomic masses of eleven other isotopes were significantly improved over previous uncertainties listed in AME2020 \cite[]{Huang_2021,Wang_2021}. The data is presently under evaluation and will be discussed in separate dedicated further publications.\par

\begin{figure}[]
\begin{center}
\includegraphics[width=0.48\textwidth, clip]{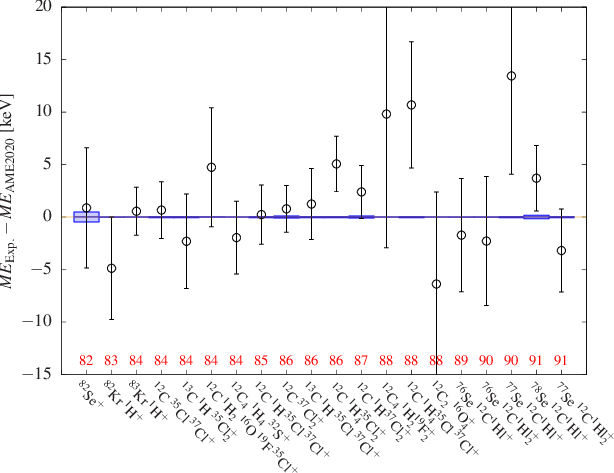}
\end{center}
	\caption{Study of the mass accuracy using singly-charged isobaric ions from time-of-flight spectra at mass numbers 82-91. Each data point denotes an independent mass measurement using a reference species of same mass number. Blue bars: uncertainty of the ion species derived from the data in AME2020. Red numbers: atomic mass number of the ion species.}
\label{fig:Chen}
\end{figure}

Various different stable ion species, well-known radioactive ion species, and molecular ions were available as isobars throughout the on-line measurements for most of the observed spectra. This rich data set allowed to make a first study the of accuracy of the mass measurements along with the observation of unknown (or not well known) radioactive ion species. In Fig.~\ref{fig:Chen} a mass evaluation of twenty different ion species with mass numbers 82-91 is shown. As presently only isobaric mass calibration is used, systematic (mass dependent) effects become small, and in this analysis only the statistical TOF uncertainties are taken into account. The isobaric reference ions used for the measurement at each mass number are (listed from lower to higher mass number): $^{82-84}$Kr$^{+}$, $^{84}$Kr$^{1}$H$^+$, $^{86}$Kr$^{+}$, $^{86}$Kr$^{1}$H$^+$, $^{12}$C$_{3}^{1}$H$_{}^{16}$O$_{2}^{19}$F$^{+}$, $^{89}$Rb$^{+}$, $^{90}$Kr$^{+}$, $^{79}$Br$_{}^{12}$C$^{+}$. Unstable ions like $^{90}$Kr$^{+}$ were part of the on-line beam.\par
For this data set of twenty different ion species, the weighted mean deviation between the measured masses and the AME2020 data was $2.26(0.80)\,\mathrm{keV/c^2}$, from which we can infer a relative mass accuracy of ${\delta m / m  = 2.80(99) \times 10^{-8}}$. 

\section{\label{Summary} Summary}
The new multi-reflection time-of-flight experiment for nuclear mass measurements downstream of the ZeroDegree spectrometer of RIKEN's RIBF facility has been introduced and relevant technical details have been discussed. An alternative in-MRTOF ion selection scheme has been introduced and was applied in on-line experiments. For the optimization procedure of the MRTOF ion mirrors, a useful method has been introduced which exploits the capability of a pulsed drift tube located in the transfer section of the system. Using the pulsed drift tube, the response of the time-of-flight for a wide energy range could be measured and provided a clear response to the modification of ion mirror voltages. This way of tuning was followed by a careful fine-tuning procedure, and enabled mass resolving powers exceeding $R_m = 1\,000\,000$, which is a milestone for MRTOF-MS and has been demonstrated using measurement accumulation times of one hour and high ion count statistics. In 2020 the new setup was coupled to a cryogenic gas cell and commissioned on-line with parasitic radioisotope beams within the HiCARI project for in-beam $\gamma$-ray spectroscopy. During this on-line commissioning period, crucial tests of the new system could be undertaken and the readiness of the setup allowed to perform atomic mass measurements of more than 70 different isotopes, where improvements of the previously known mass values have been achieved as well as newly measured masses. The setup is being improved continuously and high-precision nuclear mass measurements of exotic isotopes are planned to be performed on a regular bases for the upcoming years.

\section{\label{Acknowledgements} Acknowledgements}
We express our gratitude to the RIKEN Nishina Center for Accelerator-based Science, the Center for Nuclear Science at Tokyo University, and the HiCARI collaboration for their support of the on-line measurements. This work was supported by the Japan Society for the Promotion of Science KAKENHI (Grants No. 2200823, No. 24224008, No. 24740142, No. 15H02096, No. 15K05116, No. 17H01081, No. 17H06090, No. 18K13573, No. 19K14750, and 20H05648), and the RIKEN programme for Evolution of Matter in the Universe (r-EMU).

%

\end{document}